\documentclass{amsart}   
\usepackage{amssymb}   
\usepackage{latexsym}   
   
\newcommand{\ZZ}{{\mathbb Z}}   
\newcommand{\RR}{{\mathbb R}}   
\newcommand{\CC}{{\mathbb C}}   
\newcommand{\NN}{{\mathbb N}}   
   
\newcommand{\aA}{{\mathbb A}}

\newtheorem{theorem}{Theorem}[section]   
\newtheorem{remark}[theorem]{Remark}   
   
\newtheorem{lemma}[theorem]{Lemma}   
\newtheorem{prop}[theorem]{Proposition}   
\newtheorem{coro}[theorem]{Corollary}   
\newtheorem{definition}[theorem]{Definition}   
\newcommand{\tr}{{\mathrm{tr}}}   
   
\newcommand{\CalG}{\mathcal{G}}   
\newcommand{\CalGomega}{\mathcal{G}(\Omega,T)}   
   
\newcommand{\CalN}{\mathcal{N}(\Omega,T, \mu)}   
\newcommand{\CalA}{\mathcal{A} (\Omega,T)}   
\newcommand{\CalAf}{\mathcal{A}^{fin} (\Omega,T)}   
\newcommand{\CalKf}{\mathcal{K}^{fin} (\Omega,T)}   
\newcommand{\CalP}{\mathcal{P}}   
\newcommand{\CalX}{\mathcal{X}}   
\newcommand{\CalH}{\mathcal{H}}   
\newcommand{\CalF}{\mathcal{F}}   
\newcommand{\CalV}{\mathcal{V}}   
\newcommand{\CalD}{\mathcal{D}}   
\newcommand{\CalM}{\mathcal{M}}   
\newcommand{\CalB}{\mathcal{B}}

\newcommand{\CalY}{\mathcal{Y}}   
\newcommand{\CalE}{\mathcal{E}}   
\newcommand{\supp}{\mbox{supp}}   
\newcommand{\xo}{\mathcal{X}\times_\Omega\mathcal{X}}

\newcommand{\Oomega}{(\Omega,T)}

\newcommand{\dist}{\mbox{dist}}   
   
\begin{document}   
\title[Algebras of Random operators]{Algebras of Random operators associated   
to Delone dynamical systems}

\author[D. Lenz, P. Stollmann]{Daniel Lenz~$^{2}$,\   
Peter Stollmann~$^{3}$}

\begin{abstract} We carry out a careful study of operator algebras    
associated with Delone dynamical systems.   
A von Neumann algebra is defined using noncommutative integration theory.   
Features of these algebras and the operators they contain  are discussed.   
We restrict our attention to a certain $C^*$-subalgebra to discuss a   
Shubin trace formula.   
   
\end{abstract}   
   
\maketitle  \vspace{0.3cm}   
\noindent\footnote{Research partly supported   
by the DFG in the priority program Quasicrystals}$^{,2}$ Fakult\"{a}t f\"{u}r   
Mathematik, Technische   
Universit\"{a}t Chemnitz,   
D-09107 Chemnitz, Germany;   
E-mail:  D.Lenz@mathematik.tu-chemnitz.de,\\[0.1cm]   
   
\noindent   
$^{1,3}$  Fakult\"{a}t f\"{u}r Mathematik, Technische Universit\"{a}t   
Chemnitz,   
D-09107 Chemnitz, Germany; E-mail:  P.Stollmann@mathematik.tu-chemnitz.de

\section*{Introduction}   
The present paper is part of a study of Hamiltonians for aperiodic
solids. Among them, special emphasis is laid on models for
quasicrystals. To describe aperiodic order, we use Delone (Delaunay)
sets. Here we construct and study certain operator algebras which can be
naturally associated with Delone sets and reflect the aperiodic order
present in a Delone dynamical system. In particular, we use Connes
noncommutative integration theory to build a von Neumann algebra. This is
achieved in Section 2 after some preparatory definitions and results
gathered in Section 1. Let us stress the following facts: it is not too
hard to write down explicitely the von Neumann algebra $\CalN$ of
observables, starting from a Delone dynamical system $\Oomega$ with an
invariant measure $\mu$. As in the case of random operators, the
observables are families of operators, indexed by a set $\Omega$ of
Delone sets. This set represents a type of (aperiodic) order and the
ergodic properties of $\Oomega$ can often be expressed by combinatorial
properties of its elements $\omega$. The latter are thought of as
realizations of the type of disorder described by $\Oomega$.  The
algebra $\CalN$ incorporates this disorder and playes the role of a
noncommutative space underlying the algebra of observables. To see that
this algebra is in fact a von Neumann algebra is by no means clear. At
that point the analysis of Connes \cite{ni} enters the picture.
   
In order to verify the necessary regularity properties we rely on   
work done in \cite{oamp}, where we studied topological properties   
of a groupoid that naturally comes with $\Oomega$. Using this, we   
can construct a measurable (even topological) groupoid. Any invariant   
measure $\mu$ on the dynamical system gives rise to a transversal   
measure $\Lambda$ and the points of the Delone sets are used to define   
a random Hilbert space $\CalH$. This latter step uses specifically the   
fact that we are dealing with a dynamical system consisting of   
point sets and leads to a noncommutative random variable that has no   
analogue in the general framework of dynamical systems. We are then able   
to identify $\CalN$ as $End_\Lambda(\CalH)$. While in our approach we    
use noncommutative integration theory to verify that a certain algebra   
is a von Neumann algebra we should also like to point out that at the   
same time we provide interesting examples for the theory. Of course,   
tilings have been considered in this connection quite from the start as   
seen on the cover of \cite{cf}.  However, we emphasize the point of view   
of concrete operators and thus are led to a somewhat different setup.   
   
The study of traces on this algebra is started in Section 3. Traces are   
intimately linked to transversal functions on the groupoid. These can   
also be used to study certain spectral properties of the operator families   
constituting the von Neumann algebra. For instance, spectral properties   
are almost surely constant for the members of any such family. This type   
of results is typical for random operators. In fact, we regard the   
families studied here in this random context. An additional feature   
that is met here is the dependence of the Hilbert space on the random   
parameter $\omega\in\Omega$.    
   
In Section 4 we introduce a $C^*$-algebra that had already been encountered   
in a different form in \cite{BHZ,ke}. Our presentation here is geared towards   
using the elements of the $C^*$-algebra as tight binding hamiltonians in   
a quantum mechanical description of disordered solids (see \cite{BHZ}
for related material as well). We relate certain   
spectral properties of the members of such operator families to ergodic   
features of the underlying dynamical system. Moreover, we show that the    
eigenvalue counting functions of these operators are convergent. The limit,   
known as the integrated density of states, is an object of fundamental   
importance from the solid state physics point of view. Apart from proving its   
existence, we also relate it to the canonical trace on the von   
Neumann algebra $\CalN$ in case that the Delone dynamical system    
$\Oomega$ is  uniquely ergodic. Results of this genre are known as Shubin's   
trace formula due to the celebrated results from \cite{shu}.    

\medskip

We conclude this section with two further remarks. 

Firstly, let us mention that  starting with the work of
Kellendonk \cite{ke}, $C^\ast$-algebras associated to tilings have been
subject to intense research within the framework of K-theory (see e.g. 
\cite{ke2,kp,put}).
 This can be seen as part of a program originally initiated by
 Bellissard and his co-workers in the study of so called gap-labelling
 for almost periodic operators \cite{be1,be2,be3}. While the 
$C^\ast$-algebras we encounter
 are essentially the same, our motivation, aims and results are quite
 different.

Secondly,  let us remark that some of the results below have been   
announced in \cite{ls1,oamp}. A stronger ergodic theorem will be found   
in \cite{ls2} and a spectral theoretic application is given in \cite{kls}.

\section{Delone dynamical systems and coloured Delone dynamical systems}   
   
In this section we recall standard concepts from the theory   
of Delone sets and introduce a suitable topology on the closed sets   
in euclidian space. A slight extension concerns the discussion of   
{\it coloured (decorated) Delone sets}.   
   
\medskip   
 A subset $\omega$ of   
$\RR^d$ is called a {\it Delone set} if there exist $0<r,R<\infty$   
such that $2 r \leq \|x-y\|$ whenever $x,y\in \omega$ with $x\neq y$,   
and $B_R (x)\cap \omega \neq \emptyset$ for all $x\in \RR^d$.  Here,   
the Euclidean norm on $\RR^d$ is denoted by $\|\cdot\|$ and $B_s (x)$   
denotes the (closed) ball in $\RR^d$ around $x$ with radius $s$. The   
set $\omega$ is then also called an $(r,R)$-set.  We will sometimes be   
interested in the restrictions of Delone sets to bounded   
sets. In order to treat these restrictions, we introduce the following   
definition.

\begin{definition}{\rm (a)} A pair  $(\Lambda,Q)$ consisting of a bounded   
  subset $Q$ of $\RR^d$ and $\Lambda\subset Q$ finite is called a   
  {\rm pattern}.   
  The set $Q$ is called the {\rm support of the pattern}. \\   
  {\rm (b)} A pattern $(\Lambda,Q)$ is called a {\rm ball pattern} if   
  $Q=B_s (x)$ with $x\in \Lambda$ for suitable $x\in \RR^d$ and $s\in   
  (0,\infty)$.   
\end{definition}

The pattern $(\Lambda_1,Q_1)$ is contained in the pattern   
$(\Lambda_2,Q_2)$ written as $(\Lambda_1,Q_1) \subset (\Lambda_2,Q_2)$   
if $Q_1\subset Q_2$ and $\Lambda_1=Q_1\cap \Lambda_2$.  Diameter,   
volume etc. of a pattern are defined to be the diameter, volume etc of   
its support.  For patterns $X_1=(\Lambda_1,Q_1) $ and $X_2=   
(\Lambda_2,Q_2)$, we define $\sharp_{X_1} X_2$, {\it the number of   
  occurences of $X_1$ in $X_2$}, to be the number of elements in   
$\{t\in \RR^d : \Lambda_1 +t \subset \Lambda_2, Q_1 +t \subset Q_2\}$.   
   
For further investigation we will have to identify patterns that are   
equal up to translation. Thus, on the set of patterns we introduce an   
equivalence relation by setting $(\Lambda_1,Q_1)\sim (\Lambda_2, Q_2)$   
if and only if there exists a $t\in \RR^d$ with $\Lambda_1 = \Lambda_2   
+ t$ and $Q_1=Q_2 + t$.  In this latter case we write   
$(\Lambda_1,Q_1)= (\Lambda_2, Q_2)+ t$.  The class of a pattern   
$(\Lambda,Q)$ is denoted by $[(\Lambda,Q)]$.  The notions of diameter,   
volume, occurence etc. can easily be carried over from patterns to   
pattern classes.

Every Delone set $\omega$ gives rise to a set of pattern classes,   
$\CalP (\omega)$ viz $\CalP (\omega)=\{ [Q\wedge \omega] :   
Q\subset\RR^d\: \mbox{bounded and measurable} \}$, and to a set of   
ball pattern classes $\CalP_B (\omega)) =\{ [B_s (x)\wedge \omega] :   
x\in \omega, s>0\}$.  Here we set $Q\wedge \omega= (\omega \cap Q,   
Q)$.   
   
For $s\in (0,\infty)$, we denote by $\CalP_B^s (\omega)$ the set of   
ball patterns with radius $s$; note the relation with $s$-patches as   
considered in \cite{l1}.  A Delone set is said to be of {\it   
  finite local complexity} if for every radius $s$ the set    
$\CalP_B^s (\omega)$ is   
finite. We refer the reader to \cite{l1} for a detailed discussion of   
Delone sets of finite type.   
   
Let us now extend this framework a little, allowing for coloured Delone   
sets.   The alphabet $\aA$ is the set of possible {\it colours} or   
{\it decorations}. An $\aA$-{\it coloured Delone set} is a subset   
$\omega\subset \RR^d\times\aA$ such that the projection $pr_1(\omega)\subset   
\RR^d$ onto the first coordinate is a Delone set.   
The set of all $\aA$-coloured Delone sets is denoted by $\CalD_{\aA}$.   
   
Of course, we speak of an   
$(r,R)$-{\it set} if $pr_1(\omega)$ is an $(r,R)$-{set}. The notions of   
{\it pattern, diameter, volume} of pattern etc. easily extend to   
coloured Delone sets. E.g.   
   
\begin{definition} A pair  $(\Lambda,Q)$ consisting of a bounded             
% 1.2.   
  subset $Q$ of $\RR^d$ and $\Lambda\subset Q\times\aA$ finite is called   
  an {\rm  $\aA$- decorated pattern}.   
  The set $Q$ is called the {\rm support of the pattern}.   
\end{definition}   
   
A coloured Delone set $\omega$ is thus viewed as a Delone set $pr_1(\omega)$   
whose points $x\in pr_1(\omega)$ are labelled by colours $a\in\aA$.   
Accordingly, the translate $T_t\omega$ of a coloured Delone set $\omega   
\subset \RR^d\times\aA$ is given by   
$$   
T_t\omega=\{ (x+t,a):(x,a)\in\omega\} .   
$$   
From \cite{oamp} we infer the notion of the {\it natural topology}, defined   
on the set $\CalF (\RR^d)$ of closed subsets of $\RR^d$. Since in our   
subsequent study in \cite{ls2} the alphabet is supposed to be a   
finite set, the following   
construction will provide a suitable topology for coloured Delone sets.   
Define, for $a\in\aA$,   
$$   
p_a:\CalD_{\aA}\to \CalF (\RR^d) , p_a(\omega)=\{ x\in\RR^d: (x,a)\in   
\omega \} .   
$$   
The initial topolgy on $\CalD_{\aA}$ with respect to the family $(p_a)_   
{a\in\aA}$ is called the {\it natural topology} on the set of $\aA$-   
decorated Delone sets. It is obvious that metrizability and compactness   
properties carry over from the natural topology without decorations to   
the decorated case.   
   
Finally, the notions of Delone dynamical system and Delone dynamical   
system of finite local complexity carry over to the coloured case in    
the obvious manner.   
   
\begin{definition}                  % 1.3.   
Let $\aA$ be a finite set.   
 {\rm (a)} Let $\Omega$ be a set of Delone sets.   
  The pair $(\Omega,T)$ is called a {\em Delone dynamical system}   
  (DDS) if $\Omega$ is invariant under the shift $T$ and closed in the   
  natural   
  topology. \\   
{\rm (a')} Let $\Omega$ be a set of $\aA$-coloured Delone sets.   
  The pair $(\Omega,T)$ is called an {\em $\aA$-coloured  Delone   
   dynamical system}   
  ($\aA$-DDS) if $\Omega$ is invariant under the shift $T$ and closed in the   
  natural   
  topology. \\   
  {\rm (b)} A DDS $\Oomega$ is said to be of {\em finite local complexity} if   
  $\cup_{\omega\in \Omega} P_B^s (\omega)$ is finite for every $s>0$. \\   
{\rm (b')} An  $\aA$-DDS $\Oomega$ is said to be of {\em finite local    
  complexity} if   
  $\cup_{\omega\in \Omega} P_B^s (\omega)$ is finite for every $s>0$. \\   
 {\rm (c)} Let $0<r,R<\infty$ be given.  A DDS $\Oomega$ is said to   
  be an   
  $(r,R)$-{\em system} if every $\omega\in \Omega$ is an $(r,R)$-set.\\   
{\rm (c')} Let $0<r,R<\infty$ be given.  An  $\aA$-DDS $\Oomega$ is said to   
  be an   
  $(r,R)$-{\em system} if every $\omega\in \Omega$ is an $(r,R)$-set.\\   
  {\rm (d)} The set $\CalP (\Omega)$ of {\em pattern classes   
    associated to a DDS} $\Omega$ is defined by $\CalP   
  (\Omega)=\cup_{\omega\in \Omega} \CalP (\omega)$.   
\end{definition}   
In view of the compactness properties known for Delone sets, \cite{oamp},   
we get that $\Omega$ is compact whenever $\Oomega$ is a DDS or an $\aA$-DDS.

\section{Groupoids and non commutative random variables}    % 2.   
   
In this section we use concepts from Connes   
non-commutative integration theory \cite{ni} to associate a natural von   
Neumann algebra with a given DDS $\Oomega$. To do so, we introduce   
\begin{itemize}   
\item   
a suitable {\it groupoid} $\CalGomega$,   
   
\item   
a {\it transversal measure} $\Lambda=\Lambda_\mu$ for a given invariant   
measure $\mu$ on $\Oomega$   
   
\item and a $\Lambda$-{\it random Hilbert space} $\CalH =  (\CalH_\omega)   
_{\omega\in\Omega}$   
\end{itemize}   
leading to the von Neumann algebra   
$$   
\CalN  := \mbox{End}_\Lambda(\CalH)   
$$   
of {\it random operators}, all in the terminology of \cite{ni}. Of course, all   
these objects will now be properly defined and some crucial properties have to   
be checked. Part of the topological prerequisites have already been   
worked out in \cite{oamp}. Note that comparing the latter with the present   
paper, we put more emphasis on the relation with noncommutative   
integration theory.

The definition of the groupoid structure is straightforward see also    
\cite{BHZ},   
Sect. 2.5.   
A set $\CalG$ together with a partially defined associative   
multiplication $\cdot : \CalG^2\subset \CalG\times   
\CalG\longrightarrow \CalG$, and an inversion $^{-1}   
:\CalG\longrightarrow \CalG$ is called a groupoid if the following   
holds:   
   
\begin{itemize}   
\item $(g^{-1})^{-1}=g$ for all $g \in \CalG$,   
\item If $g_1 \cdot g_2$ and $g_2 \cdot g_3$ exist, then $g_1 \cdot   
  g_2 \cdot g_3$ exists as well,   
\item $g^{-1} \cdot g$ exists always and $g^{-1} \cdot g \cdot h = h$,   
  whenever $g \cdot h$ exists,   
\item $h \cdot h^{-1}$ exists always and $g \cdot h \cdot h^{-1} = g$,   
  whenever $g \cdot h$ exists.   
\end{itemize}   
   
A groupoid is called topological groupoid if it carries a topology   
making inversion and multiplication continuous. Here, of course,   
$\CalG\times \CalG$ carries the product topology and $\CalG^2\subset   
\CalG\times \CalG$ is equipped with the induced topology.   
   
A given groupoid $\CalG$ gives rise to some standard objects: The   
subset $\CalG^0 = \{ g \cdot g^{-1} \mid g \in \CalG \}$ is called the   
set of {\it units}. For $g \in \CalG$ we define its {\it range} $r(g)$   
by $r(g) =   
g \cdot g^{-1}$ and its {\it source} by $s(g) = g^{-1} \cdot g$. Moreover,   
we set $\CalG^\omega = r^{-1}(\{ \omega \})$ for any unit $\omega \in   
\CalG^0$. One easily checks that $g \cdot h$ exists if and only if   
$r(h) = s(g)$.   
   
By a standard construction we can assign a groupoid $\CalGomega$ to a   
Delone dynamical system.  As a set $\CalGomega$ is just $\Omega\times   
\RR^d$. The multiplication is given by $(\omega,x)   
(\omega-x,y)=(\omega, x +y)$ and the inversion is given by   
$(\omega,x)^{-1}=(\omega-x,-x)$.  The groupoid operations can be   
visualized by considering an element $(\omega,x)$ as an arrow   
$\omega-x\stackrel{x}{\longrightarrow} \omega$. Multiplication then   
corresponds to concatenation of arrows; inversion corresponds to   
reversing arrows and the set of units $\CalGomega^0$ can be identified with   
$\Omega$.   
   
Apparently this groupoid $\CalGomega$ is a topological groupoid when   
$\Omega$ is equipped with the topology of the previous section and   
$\RR^d$ carries the usual topology.   
   
The groupoid $\CalGomega$ acts naturally on a certain topological   
space $\CalX$. This space and the action of $\CalG$ on it are of   
crucial importance in the sequel. The space $\CalX$ is given by   
$$\CalX=\{(\omega,x)\in \CalG : x\in \omega\}\subset \CalGomega.$$   
In particular, it inherits a topology form $\CalGomega$.   
This $\CalX$ can be used to define a {\it random variable} or   
{\it measurable functor} in the sense of \cite{ni}. Following the latter   
reference, p. 50f, this means that we are given a functor $F$ from   
$\CalG$ to the category of measurable spaces with the following properties:   
\begin{itemize}   
\item For every $\omega\in \CalG^0$ we are given a measure space $F(\omega)=   
(\CalY^\omega,\beta^\omega)$.   
\item For every $g\in\CalG$ we have an isomorphism $F(g)$ of measure spaces,   
$F(g):\CalY^{s(g)}\to \CalY^{r(g)}$ such that $F(g_1g_2)=F(g_1)F(g_2)$,   
whenever $g_1g_2$ is defined, i.e., whenever $s(g_1)=r(g_2)$.   
\item A measurable structure on the disjoint union   
$$   
\CalY=\cup_{\omega\in\Omega}\CalY^\omega   
$$   
such that the projection $\pi: \CalY\to \Omega$ is measurable as well as   
the natural bijection of $\pi^{-1}(\omega)$ to $\CalY^\omega$.   
\item The mapping $\omega\mapsto \beta^\omega$ is measurable.   
\end{itemize}   
   
We will use the notation $F:\CalG\rightsquigarrow\CalY$ to abbreviate the   
above.   
   
Let us now turn to the groupoid $\CalGomega$ and the bundle $\CalX$ defined   
above. Since  $\CalX$ is closed (\cite{oamp}, Prop.2.1), it carries a   
reasonable Borel structure. The projection $\pi:\CalX\to\Omega$ is continuous,   
in particular measurable.   
Now, we can discuss the action of $\CalG$ on $\CalX$. Every $g =   
(\omega,x)$ gives rise to a map $J(g) : \CalX^{s(g)}\longrightarrow   
\CalX^{r(g)}$, $J(g)(\omega-x,p)= (\omega,p +x)$. A simple calculation   
shows that $J(g_1 g_2)=J(g_1) J(g_2)$ and $J(g^{-1})=J(g)^{-1}$,   
whenever $s(g_1)=r(g_2)$. Thus, $\CalX$ is an $\CalG$-space in the sense   
of \cite{lpv}. It can be used as the target space of a measurable functor   
$F:\CalG\rightsquigarrow\CalX$.   
What we still need is a {\it positive random variable} in the sense of the   
following definition, taken from \cite{oamp}. First some notation:

Given a  locally compact space $Z$, we denote the set of continuous   
functions on   
$Z$ with compact support by $C_c (Z)$. The support of a function in   
$C_c (Z)$ is denoted by $\mbox{supp}(f)$. The topology gives rise to   
the Borel-$\sigma$-algebra. The measurable nonnegative functions with   
respect to this $\sigma$-algebra will be denoted by $\CalF^+(Z)$. The   
measures on $Z$ will be denoted by $\CalM(Z)$.   
   
\begin{definition} Let $\Oomega$ be an $(r,R)$-system.    % 2.1.   
   
  {\rm (a)} A choice of measures $\beta: \Omega \to \CalM(\CalX)$ is called   
  {\rm a positive random variable with values in} $\CalX$ if the map $\omega   
  \mapsto \beta^\omega(f)$ is measurable for every $f \in   
  \CalF^+(\CalX)$, $\beta^\omega$ is supported on $\CalX^\omega$,   
  i.e., $\beta^\omega(\CalX - \CalX^\omega) = 0$, $\omega\in \Omega$,   
  and $\beta$ satisfies the following invariance condition   
\[ \int_{\CalX^{s(g)}} f(J(g)p) d\beta^{s(g)}(p) = \int_{\CalX^{r(g)}}   
f(q) d\beta^{r(g)}(q) \] for all $g \in \CalG$ and $f \in   
\CalF^+(\CalX^{r(g)})$.   
   
{\rm (b)} A map $\Omega\times C_c (\CalX)\longrightarrow \CC$ is called a   
{\rm complex random variable} if there exist an $n\in \NN$, positive random   
variables $\beta_i$, $i=1,\ldots,n$ and $\lambda_i\in\CC$,   
$i=1,\ldots,n$ with $\beta^{\omega}(f)=\sum_{i=1}^n \beta^\omega_i   
(f)$.   
\end{definition}   
   
We are now heading towards introducing and studying a special random   
variable. This variable is quite important as it gives rise to the   
$\ell^2$-spaces on which the Hamiltonians act. Later we will see that   
these Hamiltonians also induce random variables.   
   
\begin{prop} \label{alpha} Let $\Oomega$ be an $(r,R)$-system. Then the   
map  $\alpha : \Omega \longrightarrow \CalM (\CalX)$, $\alpha^{\omega} (f)   
= \sum_{p\in \omega} f(p)$ is a random variable with values in $\CalX$.   
Thus the functor $F_\alpha$ given by $F_\alpha(\omega)=(\CalX^\omega,   
\alpha^\omega)$ and $ F_\alpha(g)=J(g)$ is measurable.   
\end{prop}   
\begin{proof} See \cite{oamp}, Corollary 2.6.\end{proof}   
   
Clearly, the condition that $\Oomega$ is an $(r,R)$-system is used to   
verify the measurability conditions needed for a random variable.   
We should like to stress the fact that the above functor given by $\CalX$   
and $\alpha^\bullet$ differs from the canonical choice, possible for any   
dynamical system. In the special case at hand this canonical choice reads as follows:   
   
\begin{prop} \label{nu} Let $\Oomega$ be a DDS. Then the   
map  $\nu : \Omega \longrightarrow \CalM (\CalG)$, $\nu^{\omega} (f)   
= \int_{\RR^d} f(\omega,t)dt$ is a {\em transversal function}, i.e., a random   
variable with values in $\CalG$.   
\end{prop}   
   
Actually, one should possibly define transversal functions before introducing   
random variables. Our choice to do otherwise is to underline the specific   
functor used in our discussion of Delone sets. As already mentioned above,   
the analogue of the transversal function $\nu$ from Proposition \ref{nu}   
can be defined for any dynamical system. In fact this structure has been   
considered by Bellissard and coworkers in a $C^*$-context. The notion   
{\it almost random operators} has been coined for that; see   
\cite{be1} and the literature quoted there.   
   
After having encountered functors from $\CalG$ to the category of   
measurable spaces under the header random variable or measurable functor, we   
will now meet {\it random Hilbert spaces}. By that one designates,   
according to \cite{ni}, a {\it representation} of $\CalG$ in the category of   
Hilbert spaces, given by the following data:   
\begin{itemize}   
\item   
A measurable family  $\CalH = (\CalH_\omega)_{\omega\in \CalG^0}$ of   
Hilbert spaces.   
   
\item   
For every $g\in\CalG$ a unitary $U_g:\CalH_{s(g)}\to\CalH_{r(g)}$ such   
that   
$$ U(g_1g_2)=U(g_1)U(g_2)$$   
whenever $s(g_1)=r(g_2)$. Moreover, we assume that for   
every pair $(\xi,\eta)$ of   
measurable sections of $\CalH$ the function   
$$   
\CalG\to \CC, g\mapsto (\xi|\eta)(g):=(\xi_{r(g)}|U(g)\eta_{s(g)})   
$$   
is measurable.   
\end{itemize}   
Given a measurable functor $F:\CalG\rightsquigarrow\CalY$ there is a natural   
representation $L^2\circ F$, where   
$$   
\CalH_\omega=L^2(\CalY^\omega,\beta^\omega)   
$$   
and $U(g)$ is induced by the isomorphism $F(g)$ of measure spaces.   
   
Let us assume that $\Oomega$ is an $(r,R)$-system. We are   
especially interested in the representation of $\CalGomega$ on   
$\CalH=(\ell^2   
  (\CalX^\omega,\alpha^\omega))_{\omega\in\Omega}$   
induced by the measurable functor $F_\alpha: \CalGomega\rightsquigarrow   
\CalX$ defined above. The necessary measurable structure is provided by   
\cite{oamp}, Proposition 2.8.   It is the measurable structure generated by $C_c (\CalX)$. 
   
The last item we have to define is a {\it transversal measure}. We denote   
the set of nonnegative transversal functions on a   
groupoid $\CalG$ by $\CalE^+(\CalG)$ and consider the unimodular case   
($\delta\equiv 1$) only. Following \cite{ni}, p. 41f, a {\it transversal   
measure} $\Lambda$ is a linear mapping   
$$   
\Lambda: \CalE^+(\CalG)\to [0,\infty]   
$$   
satisfying   
\begin{itemize}   
\item   
 $\Lambda$ is {\it normal}, i.e., $\Lambda(\sup \nu_n)=\sup   
\Lambda(\nu_n)$   
for every increasing sequence $(\nu_n)$ in $\CalE^+(\CalG)$.   
\item   
 $\Lambda$ is {\it invariant}, i.e., for every  $\nu\in\CalE^+(\CalG)$ and   
every kernel $\lambda$ with $\lambda^\omega(1)=1$ we get   
$$   
\Lambda(\nu * \lambda)=\Lambda(\nu) .   
$$   
\end{itemize}   
Given a fixed transversal function $\nu$ on $\CalG$ and an invariant measure   
$\mu$ on $\CalG^0$ there is a unique transversal measure $\Lambda=\Lambda_\nu$   
such that   
$$   
\Lambda(\nu * \lambda)=\mu(\lambda^\bullet(1)) ,   
$$   
see \cite{ni}, Theoreme 3, p.43. In the next Section we will discuss that   
in a little more detail in the case of DDS groupoids.   
   
We can now put these constructions together.

\begin{definition} \label{def24} Let $\Oomega$ be an $(r,R)$-system and    
let $\mu$ be an      
% 2.4.   
invariant measure on $\Omega$. Denote by   
$\CalV_1$ the set of all $f : \CalX\longrightarrow \CC   
  $ which are measurable and satisfy $ f(\omega,\cdot)\in \ell^2   
  (\CalX^\omega,\alpha^\omega)$ for every $ \omega\in \Omega$.   
   
A  family   
$(A_\omega)_{\omega\in\Omega}$ of bounded operators $A_\omega :   
\ell^2(\omega,\alpha^\omega)\longrightarrow   
\ell^2(\omega,\alpha^\omega)$ is called {\it measurable} if   
$\omega\mapsto \langle f(\omega), (A_\omega g)(\omega)\rangle_\omega$   
is {\rm measurable} for all $f,g\in \CalV_1$. It is called {\rm bounded} if the   
norms of the $A_\omega$ are uniformly bounded. It is called {\rm covariant}   
if it satisfies the covariance condition   
$$   
H_{\omega+t} = U_t H_\omega U_t^*, \;\omega\in \Omega, t\in \RR^d,   
$$   
where $U_t :\ell^2(\omega)\longrightarrow \ell^2 (\omega + t)$ is the   
unitary operator induced by translation. Now, we can define   
$$   
\CalN:=\{ A=(A_\omega)_{\omega\in\Omega}|A\mbox{  covariant,   
measurable and bounded}\}/\sim,   
$$   
where $\sim$ means that we identify families which agree $\mu$ almost   
everywhere.   
\end{definition}   
   
As is clear from the definition, the elements of $\CalN$ are classes   
of families of operators. However, we will not distinguish too   
pedantically between classes and their representatives in the sequel.   
   
\begin{remark}{\rm  It is possible to define $\CalN$ by requiring seemingly   
weaker conditions. Namely, one can consider families $(A_\omega)$ that are   
essentially bounded and satisfy the covariance condition almost   
everywhere.   
However, by standard procedures (see \cite{ni,le}), it is possible to show   
that   
each of these families agrees almost everywhere with a family satisfying the   
stronger conditions discussed above.   
   
Obviously, $\CalN$ depends on the measure class of $\mu$ only. Hence, for   
uniquely ergodic $\Oomega$, $\CalN=:\mathcal{N}(\Omega,T)$ gives a canonical   
algebra. This case has been considered in \cite{ls1,oamp}.}   
\end{remark}   
   
Apparently, $\CalN$ is an involutive algebra under the obvious   
operations.   Moreover, it can be related to the algebra $\mbox{{\rm End}}_\Lambda(\CalH) $
defined in \cite{ni} as follows.

\begin{theorem} Let $\Oomega$ be an $(r,R)$-system and let $\mu$ be an   
invariant measure on $\Omega$. Then $\CalN$ is a weak-*-algebra. More   
precisely,   
$$   
\CalN  = \mbox{{\rm End}}_\Lambda(\CalH) ,   
$$   
where $\Lambda=\Lambda_\nu$ and $\CalH=(\ell^2   
  (\CalX^\omega,\alpha^\omega))_{\omega\in\Omega}$   
are defined as above.   
\end{theorem}   
\begin{proof}   
The asserted equation follows by plugging in the respective   
definitions. The only thing that remains to be checked is that   
$\CalH$ is a square integrable representation in the sense of   
\cite{ni}, Definition, p. 80. In order to see this it suffices   
to show that the functor $F_\alpha$ giving rise to $\CalH$ is proper.   
See \cite{ni}, Proposition 12, p. 81.   
   
This in turn follows by considering the transversal function $\nu$ defined   
in Proposition \ref{nu} above. In fact, any $u\in C_c(\RR^d)^+$ gives rise   
to the function $f\in\CalF^+(\CalX)$ by $f(\omega,p):=u(p)$.   
It follows that   
$$   
(\nu*f)(\omega,p)=\int_{\RR^d}u(p+t)dt=\int_{\RR^d}u(t)dt ,   
$$   
so that $\nu*f\equiv 1$ if the latter integral equals $1$ as required by   
\cite{ni}, Definition 3, p. 55.   
\end{proof}   
We can use the measurable structure to identify $L^2 (\CalX,m)$, where   
$m=\int_\Omega \alpha^\omega\mu(\omega)$   
with $\int_\Omega^{\oplus} \ell^2   
(\CalX^\omega,\alpha^\omega)\,d\mu(\omega)$. This gives the faithful   
representation   
$$   
\pi:   
\CalN\longrightarrow B(L^2 (\CalX,m)),   
\pi(A)   
f((\omega,x))= (A_\omega f_\omega)((\omega,x))   
$$   
and the following immediate consequence.   
\begin{coro}  $\pi (\CalN)\subset B(L^2 (\CalX,m)) $   
is a von Neumann algebra.   
\end{coro}

Next we want to identify conditions under which $\pi (\CalN)$ is a factor.   
Recall that a Delone set $\omega$ is said to be {\it non-periodic} if   
$\omega+t=\omega$ implies that $t=0$.

\begin{theorem} Let $\Oomega$ be an $(r,R)$-system and let $\mu$ be an   
ergodic invariant measure on $\Omega$. If   $\omega$ is non-periodic   
for $\mu$-a.e. $\omega\in\Omega$ then $\CalN$ is a factor.   
\end{theorem}   
\begin{proof}   
We want to use \cite{ni}, Corollaire 7, p. 90.    In our case $\CalG=\CalGomega$, $\CalG^0=\Omega$ and   
$$   
\CalG^\omega_\omega=\{ (\omega,t):\omega+t=\omega \} .   
$$   
Obviously, the latter is trivial, i.e., equals $\{ (\omega,0)\}$ iff   
 $\omega$ is non-periodic. By our assumption this is valid $\mu$-a.s.   
so that we can apply \cite{ni}, Corollaire 7, p. 90.   
   Therefore the centre of $\CalN$ consists of families   
$$   
f=(f(\omega)1_{\CalH_\omega})_{\omega\in\Omega},   
$$   
where $f:\Omega\to \CC$ is bounded, measurable and invariant. Since $\mu$   
is assumed to be ergodic this implies that $f(\omega)$ is a.s. constant so   
that the centre of $\CalN$ is trivial.   
\end{proof}   
   
\begin{remark}{\rm  Since $\mu$ is  ergodic, the assumption of non-periodicity in   
the theorem can be replaced by assuming that there is a set of positive   
measure consisting of non-periodic $\omega$.   
   
\medskip

Note that the latter result gives an extension of part of what has been   
announced in \cite{ls1}, Theorem 2.1 and \cite{oamp}, Theorem 3.8. The   
remaining assertions of \cite{oamp} will be proven in the following Section, again in   
greater generality.}   
\end{remark}   

\section{Transversal functions, traces and deterministic spectral properties.}   
   
In the preceding section we have defined the von Neumann algebra $\CalN$   
starting from an $(r,R)$-system $\Oomega$ and an invariant measure $\mu$ on   
$\Oomega$. In the present section we will study traces on this algebra.   
Interestingly, this rather abstract and algebraic enterprise will lead to   
interesting spectral consequences. We will see that the operators involved   
share some fundamental properties with ``usual random operators''.   
   
Let us first draw the connection of our families to ``usual random operators'',   
referring to \cite{cl,pf,s} for a systematic account. Generally   
speaking one is concerned with families $(A_\omega)_{\omega\in\Omega}$   
of operators   
indexed by some probability space and  acting on $\ell^2(\ZZ^d)$ or   
$L^2(\RR^d)$ typically. The probability space $\Omega$ encodes some   
statistical properties, a certain kind of disorder that is inspired by   
physics in many situations. One can view the set $\Omega$ as the set of   
all possible realization of a fixed disordered model and each single   
$\omega$ as a possible realization of the disorder described by $\Omega$. Of   
course, the information is mostly encoded in a measure on   
$\Omega$ that describes the probability with which a certain realization is   
picked.   
   
We are faced with a similar situation, one difference being that in any   
family $A=(A_\omega)_{\omega\in\Omega}\in\CalN$, the operators $A_\omega$   
act on the possibly different spaces $\ell^2(\omega)$. Apart from that   
we have the same ingredients as in the usual random business, where, of   
course, Delone dynamical systems still bear quite some order. That is,   
we are in the realm of weakly disordered systems. For a first idea what this   
might have to do with aperiodically ordered solids, quasicrystals, assume   
that the points $p\in\omega$ are the atomic positions of a quasicrystal. In a   
tight binding approach (see \cite{BHZ} Section 4 for why this is reasonable),   
the Hamiltonian $H_\omega$ describing the respective solid would naturally   
be defined on $\ell^2(\omega)$, its matrix elements $H_\omega(p,q),   
p,q\in\omega$ describing  the diagonal and hopping terms for an   
electron that undergoes the influence of the atomic constellation given   
by $\omega$. The definite choice of these matrix elements has to be   
done on physical grounds. In the following subsection we will propose   
a $C^*$-subalgebra that contains what we consider the most reasonable   
candidates; see also \cite{BHZ, ke}.   
It is clear, however, that $\CalN$ is a reasonable framework,   
since translations should not matter. Put in other words, every reasonable   
Hamiltonian family $(H_\omega)_{\omega\in\Omega}$ should be covariant.   
   
The remarkable property that follows from this ``algebraic'' fact is that   
certain spectral properties of the $H_\omega$ are {\it deterministic}, i.e.,   
do not depend on the choice of the realization $\omega$ $\mu$-a.s.   
   
Let us next introduce the necessary algebraic concepts, taking a second   
look at transversal functions and random variables with values in
$\CalX$.  In fact, random variables can be integrated with respect to
transversal measures by \cite{ni}, i.e  for a given non-negative random variable $\beta$
with values in $\CalX$ and a transversal measure $\Lambda$, the
expression   $\int F_\beta d\Lambda $ is well defined. 
  More precisely, the following holds:

\begin{lemma}\label{Integration} Let $\Oomega$ be an $(r,R)$-system and   
$\mu$ be  $T$-invariant.   
   
{\rm (a)} Let $\beta$ be a nonnegative random variable   
with values in $\CalX$. Then $\int_\Omega   
  \beta^{\omega} (f(\omega,\cdot))\,d\mu(\omega)$ does not depend on   
  $f \in \CalF^+ (\CalX)$ provided $f$ satisfies $\int f((\omega+t,   
  x+t )\,dt=1$ for every $(\omega,x)\in\CalX$ and   
$$   
\int_\Omega   
  \beta^{\omega} (f(\omega,\cdot))\,d\mu(\omega)=\int F_\beta d\Lambda ,   
$$   
where $F_\beta:\CalG \rightsquigarrow \CalX$ is the measurable functor   
induced by $F_\beta(\omega)=(\CalX^\omega,\beta^\omega)$ and $\Lambda=   
\Lambda_\nu$ the transversal measure defined in the previous section.   
   
{\rm (b)} An analogous statement remains true for a complex random   
variable $\beta=\sum_k\lambda_k\beta_k$, when we define   
$$\int F_\beta d\Lambda =\sum_k \lambda_k \int F_{\beta_k} d\Lambda   
$$   
and restrict to $f \in \CalF^+ (\CalX)$ with ${\rm supp} f$ compact.   
\end{lemma}   
\begin{proof} Part (a) is a direct consequence of the definitions   and
 results in \cite{ni}. Part (b), then easily follows from (a) by
 linearity. 
\end{proof}
 
A special instance of the foregoing lemma is given in the following
proposition.

\begin{prop}\label{infinity}   
Let $(\Omega, T)$ be an $(r, R)$-system and let $\mu$ be $T$-invariant. If   
$\lambda$ is a transversal function on $G (\Omega, T)$ then   
$$   
\varphi\mapsto\int_\Omega\langle \lambda^\omega,\varphi\rangle d\mu(\omega)   
$$   
defines an invariant functional on $C_c (\RR^d)$, i.e., a multiple of the   
Lebesgue measure.   
In particular, if $\mu$ is an ergodic measure, then  either $\lambda^{\omega} (1) = 0$ a.s. or   
$\lambda^{\omega} (1) = \infty$ a.s.   
\end{prop}   
\begin{proof}  Invariance of the functional follows by direct
  checking. By    
uniqueness of the Haar measure, this functional must then be a multiple
of Lebesgue measure. If $\mu$ is ergodic, the map $\omega \mapsto
\lambda^{\omega} (1) $ is almost surely constant (as it is obviously
invariant). This easily implies the last statement.     
\end{proof}   

Each random operator gives rise to a random variable as seen in the   
following proposition whose simple proof we omit.   
   
\begin{prop}  Let $\Oomega$ be an $(r,R)$-system and   
$\mu$ be  $T$-invariant.   
Let $(A_\omega)\in \CalN$ be given. Then the map $\beta_A :   
\Omega \longrightarrow \CalM(\CalX)$, $\beta_A ^\omega(f)=\tr (A_\omega   
M_f)$ is a complex random variable with values in $\CalX$.   
\end{prop}

Now, choose a nonnegative  measurable $u$ on  $\RR^d$ with  compact
support and $\int_{\RR^d} u(x)   
dx =1$.  Combining the previous proposition with Lemma   
\ref{Integration}, $f(\omega,p):=u(p)$, we infer that the map   
$$\tau : \CalN\longrightarrow \CC, \;\: \tau(A)=\int_\Omega   
\tr(A_\omega M_u)\,d\mu(\omega)$$   
does not depend on the choice of   
$f$ viz $u$ as long as the integral is one. Important features of    
$\tau$ are given in the following lemma.   
   
\begin{lemma}  Let $\Oomega$ be an $(r,R)$-system and   
$\mu$ be  $T$-invariant.   
Then the map $\tau:\CalN\longrightarrow \CC$ is continuous,   
faithful,   
nonegative on $\CalN^+$ and satisfies $\tau(A) = \tau(U^*AU)$   
for every unitary $U\in\CalN$ and arbitrary $A\in\CalN$, i.e., $\tau$   
is a trace.   
\end{lemma}   
%%%%%%%%%%%%%%%%%%%%%   
We include the elementary proof, stressing the fact that we needn't rely   
on the noncommutative framework; see also \cite{lpv} for the respective    
statement in a different setting.   
%%%%   
\begin{proof}   Choosing a continuous $u$ with compact support we 
  see that $|\tau(A) - \tau(B)|\leq \int \|A_\omega - B_\omega\| \tr M_u
  d\mu(\omega)\leq \| A -  B\| C$, where $C>0$ only depends on $u$ and
  $\Omega$. On the other hand, choosing $u$ with arbitrary large support
  we easily infer that $\tau$ is faithful. It remains to show the last
  statement. 

According to \cite{dix}, I.6.1, Cor.1 it suffices to show    
$\tau(K^*K)=\tau(KK^*)$   
for every $K=(K_\omega)_{\omega\in\Omega}\in\CalN$. We write   
$k_\omega(p,q):=(K_\omega\delta_q|\delta_p)$ for the associated   
kernel and calculate   
\begin{eqnarray*}   
\tau(K^*K)&=&\int_\Omega \tr(K_\omega^*K_\omega M_u)d\mu(\omega)\\   
&=&\int_\Omega \tr(M_{u^\frac12}K_\omega^*K_\omega M_{u^\frac12})d\mu(\omega)\\   
%&=&\int_\Omega \sum_{m\in\omega}(M_{u^\frac12}K_\omega^*K_\omega    
%M_{u^\frac12}\delta_m|\delta_m)d\mu(\omega)\\   
&=&   
\int_\Omega \sum_{m\in\omega}\| K_\omega    
M_{u^\frac12}\delta_m\|^2\mu(\omega)\\   
%&=&\int_\Omega \sum_{l,m\in\omega}|k_\omega(l,m)|^2u(m)d\mu(\omega)\\   
&=&\int_\Omega \sum_{l,m\in\omega}|k_\omega(l,m)|^2u(m)\int_{\RR^d}   
u(l-t)dtd\mu(\omega)   
\end{eqnarray*}   
where we used that $\int_{\RR^d}   
u(l-t)dt=1$ for all $l\in\omega$.
 By covariance and Fubinis theorem we get   
$$\cdots = \int_{\RR^d} \int_\Omega \sum_{l,m\in \omega} |k_{\omega-t} (l-t,m-t)|^2u(m)   
u(l-t) d\mu(\omega) dt.$$
As $\mu$ is $T$-invariant, we can replace $\omega-t$ by $\omega$ and obtain
\begin{eqnarray*}   
&=&  \int_{\RR^d}  \int_\Omega \sum_{l,m\in\omega  + t }|k_\omega(l-t,m-t)|^2u(m)   
u(l-t)dtd\mu(\omega)\\   
&=&\int_\Omega \int_{\RR^d}\sum_{l,m\in\omega}|k_\omega(l,m)|^2u(m+t)   
u(l)dtd\mu(\omega)\\   
%&=&\int_\Omega \sum_{l,m\in\omega}|k_\omega(l,m)|^2   
%u(l)d\mu(\omega)\\   
&=&\int_\Omega \tr(K_\omega K_\omega^* M_u)d\mu(\omega)   
\end{eqnarray*}   
by reversing the first steps.   
\end{proof}   
   
Having defined $\tau$, we can now associate a canonial measure   
$\rho_A$ to every selfadjoint $A\in \CalN$.   
   
\begin{definition}\label{rho} For $A\in \CalN$ selfadjoint,  and $B\subset \RR$ Borel   
measurable,   
we set   $\rho_A(B)\equiv \tau(\chi_B(A))$, where $\chi_B$ is the   
characteristic   
function of $B$.   
\end{definition}   
For the next two results we refer to \cite{lpv} where the context is   
somewhat different.   
\begin{lemma}  Let $\Oomega$ be an $(r,R)$-system and   
$\mu$ be  $T$-invariant.   
Let $A\in \CalN$ selfadjoint be given. Then  $\rho_A$ is a   
spectral measure   
for $A$. In particular, the support of $\rho_A$ agrees with the   
spectrum   
$\Sigma$ of $A$ and the equality $\rho_A (F)=\tau (F(A))$ holds for every   
bounded   
measurable $F$ on $\RR$.   
\end{lemma}

\begin{lemma}  Let $\Oomega$ be an $(r,R)$-system and   
$\mu$ be  $T$-invariant.   
Let $\mu$ be ergodic and $A=(A_\omega)\in \CalN$ be selfadjoint.   
Then there exists $\Sigma, \Sigma_{ac},\Sigma_{sc},\Sigma_{pp},\Sigma_{ess}   
\subset \RR$ and a subset  $\widetilde{\Omega}$ of  $\Omega$ of full measure   
such that $\Sigma=\sigma(A_\omega)$ and  $\sigma_{\bullet} (A_\omega)=   
\Sigma_{\bullet}$ for $\bullet=ac,sc,pp,ess$ and $\sigma_{disc}   
(A_\omega)=\emptyset$  for every $\omega\in \widetilde{\Omega}$.  In
this case, the spectrum of $A$ is   given by $\Sigma$.  
\end{lemma}

We now head towards evaluating the trace $\tau$.   
   
\begin{definition}   
The number   
$$   
\int F_\alpha d\Lambda =: D_{\Omega,\mu}   
$$   
is called the {\em mean density} of $\Omega$ with respect to $\mu$.   
\end{definition}   
   
\begin{theorem}   
 Let $\Oomega$ be an $(r,R)$-system and   
$\mu$ be  ergodic.   If   $\omega$ is non-periodic   
for $\mu$-a.e. $\omega\in\Omega$ then $\CalN$ is a factor of type II$_D$,   
where $D=D_{\Omega,\mu}$, i.e., a finite factor of type II and the   
canonical trace $\tau$ satisfies $\tau(1)=D$.   
\end{theorem}   
   
\begin{proof}   
We already know that $\CalN$ is a factor. Using Proposition \ref{infinity}   
and \cite{ni}, Cor. 9, p. 51 we see that $\CalN$ is not of type I. Since    
it admits a finite faithful trace, $\CalN$ has to be a finite factor   
of type II.   
   
Note that Lemma \ref{Integration},   
the definition of $\tau$ and $\alpha$ give the asserted value for $\tau(1)$.   
\end{proof}   
   
\begin{remark}{\rm    It is a simple consequence of Proposition
    \ref{prop46} below that 
$$   
  D_\omega=\lim_{R\to\infty}\frac{\#(\omega\cap B_R(0))}{|B_R(0)|}   
$$   
exists  and equals $D_{\Omega,\mu}$ for almost every $\omega \in
\Omega$. Therefore, the preceding result is a more   
general version of the results announced as \cite{ls1}, Theorem 2.1 and   
\cite{oamp}, Theorem 3.8, respectively. Of course, existence of the
limit is not new. It can already be found e.g. in \cite{BHZ}. }   

\end{remark}   

\section{The C$^*$-algebra associated to  finite range operators and the integrated   
density of states}   
   
In this section we study a C$^*$-subalgebra of $\CalN$ that contains those   
operators that might be used as hamiltonians for quasicrystals. The approach   
is direct and does not rely upon the framework introduced in the   
preceding sections.   
   
We define   
$$   
\xo:=\{ (p,\omega, q)\in\RR^d\times\Omega\times\RR^d: p,q\in\omega\} ,   
$$   
which is a closed subspace of $\RR^d\times\Omega\times\RR^d$ for any DDS   
$\Omega$.   
   
\begin{definition}   
A {\rm kernel of finite range} is a function $k\in C(\xo )$ that satisfies   
the following properties:   
\begin{itemize}   
\item[{\rm (i)}] $k$ is bounded.   
\item[{\rm (ii)}] $k$ has finite range, i.e., there exists $R_k>0$ such that   
$k(p,\omega, q)=0$, whenever $|p-q|\ge R_k$.   
\item[{\rm (iii)}] $k$ is invariant, i.e.,    
$$   
k(p+t,\omega +t,q+t)=k(p,\omega,q),   
$$   
for $(p,\omega,q)\in\xo$ and $t\in\RR^d$.   
\end{itemize}   
The set of these kernels is denoted by $\CalKf$.   
\end{definition}   
%%%%%%%%%%%%%   
We record a few quite elementary observations.    
For any kernel $k \in \CalKf$ denote by $\pi_\omega
  k:=K_\omega$ the operator $K_\omega\in\CalB(\ell^2(\omega))$,
  induced by
  $$
  (K_\omega\delta_q|\delta_p):= k(p,\omega,q)\mbox{ for
    }p,q\in\omega .
  $$
  Clearly, the family $K:=\pi k$, $K=(K_\omega)_{\omega\in\Omega}$,
  is bounded in the product (equipped with the supremum norm)
  $\Pi_{\omega\in\Omega}\CalB(\ell^2(\omega))$. Now,  pointwise sum, the convolution (matrix) product
  $$
  (a\cdot b)(p,\omega,q) := \sum_{x\in\omega}
  a(p,\omega,x)b(x,\omega,q)
  $$
  and the involution $ k^*(p,\omega,q):= \overline{k}(q,\omega,p)$
  make $\CalKf$ into a $*$-algebra. Then,  the mapping $\pi:\CalKf\to\Pi_{\omega\in\Omega}
  \CalB(\ell^2(\omega))$  is a faithful
  $*$-representation.  We denote $\CalAf:=\pi(\CalKf)$ and call it the
  {\it operators of finite range}. The completion of $\CalAf$ with respect to the norm
$\| A\|:=\sup_{\omega\in\Omega}\| A_\omega\|$ is denoted  by $\CalA$. It
is not hard to see that the mapping $\pi_\omega:\CalAf\to\CalB(\ell^2(\omega)),
  K\mapsto K_\omega$  is a representation
  that extends by continuity to a representation of $\CalA$ that we
  denote by the same symbol.

\begin{prop} \label{bas} Let $A\in \CalA$ be given. Then the following holds:\\
(a) $\pi_{\omega + t}  (A) = U_t \pi_\omega (A) U_t^\ast$ for arbitrary
$\omega\in \Omega$ and $t\in \RR^d$. \\
(b) For $F \in  C_c (\CalX)$, the map $\omega \mapsto  \langle
\pi_\omega (A) F_\omega, F_\omega\rangle_\omega$ is continuous.
\end{prop}
\begin{proof} Both statements are immediate for $A\in \CalAf$ and then
  can be extended to $\CalA$ by density and the definition of the norm. 
\end{proof}

We get the following result that relates ergodicity properties of $\Oomega$,   
spectral properties of the operator families from $\CalA$ and properties   
of the representations $\pi_\omega$.   
%%%   
\begin{theorem}The following conditions on a DDS $\Oomega$ are equivalent:     
\begin{itemize}     
\item[{\rm (i)}] $(\Omega,T)$ is minimal.     
\item[{\rm (ii)}] For any selfadjoint $A\in\CalA$ the spectrum     
$\sigma(A_\omega)$ is independent of $\omega\in\Omega$.     
\item[{\rm (iii)}] $\pi_\omega$ is faithful for every $\omega\in\Omega$.     
\end{itemize}     
\end{theorem}     
\begin{proof}   
(i)$\Longrightarrow$(ii):\\   
\noindent  Choose  $\phi\in C(\RR)$. We then get $\pi_\omega(\phi(A))=   
\phi(\pi_\omega(A))$ since $\pi_\omega$ is a continuous algebra   
homomorphism.   Set $\Omega_0 = \{\omega \in \Omega :
\pi_\omega(\phi(A))=0\}$. By Proposition \ref{bas} (a), $\Omega_0$  is
invariant under translations. Moreover, by Proposition \ref {bas} (b)
it is closed. Thus, $\Omega_0 = \emptyset$ or $\Omega_0 = \Omega$ by
minimality. 
As $\phi$ is arbitrary, this gives the desired equality of spectra by spectral calculus.

(ii)$\Longrightarrow$(iii):\\   
\noindent By (ii) we get that $\|\pi_\omega(A)\|^2=\|\pi_\omega(A^*A)\|$   
does not depend on $\omega\in\Omega$. Thus $\pi_\omega(A)=0$ for some   
$A$ implies that $\pi_\omega(A)=0$ for all $\omega\in\Omega$ whence $A=0$.   
    
(iii)$\Longrightarrow$(i):\\   
Assume that $\Omega$ is not minimal. Then we find $\omega_0$ and   
$\omega_1$ such that $\omega_1\not\in\overline{(\omega_0+\RR^d)}$.   
   
Consequently , there is $r>0, p\in\omega, \delta>0$ such that   
$$   
d_H((\omega_0-p)\cap B_r(0),(\omega_1-q)\cap B_r(0))>2\delta   
$$   
for all $q\in\omega_1$. Let $\rho\in C(\RR)$ such that $\rho(t)=0$ if   
$t\ge \frac12$ and $\rho(0)=1$. Moreover, let $\psi\in C_c(\RR^d)$   
such that $\supp\psi\subset B_\delta(0)$ and $\phi\in C_c(\RR^d)$   
and $\phi=1$ on $B_{2r}(0)$.   
   
Finally, let   
\begin{eqnarray*}   
a(x,\omega,y)&:=&\rho\left( \| \left( \sum_{p\in\omega}T_p\psi\right) T_x\phi   
-\left(\sum_{q\in\omega_0}T_q\psi\right) T_y\phi\|_\infty  \right. \\   
&+&   
\left. \| \left( \sum_{p\in\omega_0}T_p\psi\right) T_x\phi   
-\left(\sum_{q\in\omega}T_q\psi\right) T_y\phi\|_\infty\right)   
\end{eqnarray*}   
It is clear that $a$ is a symmetric kernel of finite range and by   
construction the corresponding operator family satisfies   
$A_{\omega_1}=0$ but $A_{\omega_0}\not=0$, which implies (iii).   
\end{proof}   
   
Let us now comment on the relation between the algebra $\CalA$ defined above   
and the C$^*$-algebra introduced in \cite{BHZ,ke} for a different purpose    
and in a different setting. Using the notation from \cite{BHZ} we    
let   
$$   
\CalY =\{\omega\in\Omega :0\in\omega\}   
$$   
and    
$$   
G_\CalY=\{ (\omega,t)\in \CalY\times\RR^d: t\in\omega\}\subset\CalX .   
$$   
In \cite{BHZ} the authors introduce the algebra $C^*(G_\CalY)$, the   
completion of $C_c(G_\CalY)$ with respect to the    
convolution   
$$   
fg(\omega,q)=\sum_{t\in\omega}f(\omega,t)g(\omega-t,q-t)   
$$   
and the norm induced by the representations   
$$   
\Pi_\omega :C_c(G_\CalY)\to\CalB (\ell^2(\omega)), \Pi_\omega(f)\xi(q)   
=\sum_{t\in\omega}f(\omega-t,t-q)\xi(q), q\in\omega .   
$$   
The following result can be checked readily, using the definitions.   
\begin{prop}   
For a kernel $k\in\CalKf$ denote    
$f_k(\omega,t):=k(0,\omega,t)$. Then   
$$   
J:\CalKf\to C_c(G_\CalY), k\mapsto f_k   
$$   
is a bijective algebra isomorphism and $\pi_\omega=\Pi_\omega\circ J$   
for all $\omega$. Consequently, $\CalA$ and $C^*(G_\CalY)$ are isomorphic.   
\end{prop}   
Note that the setting in \cite{BHZ} and here are somewhat different. In the   
tiling framework, the analogue of these algebras have been considered in   
\cite{ke}.   
   
We now come to relate the abstract trace $\tau$ defined in the last section   
with the mean trace per unit volume. The latter object is quite often   
considered by physicists and bears the name {\it integrated density   
of states}. Its proper definition rests on ergodicity. We start with   
the following preparatory result for which we need the notion of a van Hove   
sequence of sets.   
   
For $s>0$ and $Q\subset \RR^d$, we denote by $\partial_s Q$ the set of     
points in $\RR^d$ whose distance to the boundary of $Q$ is less than     
$s$. A sequence $(Q_n)$ of bounded subsets of $\RR^d$ is called a {\it van     
Hove sequence} if $|Q_n|^{-1} |\partial_s Q_n|\longrightarrow 0,     
n\longrightarrow 0$ for every $s>0$.     
     
\begin{prop}\label{prop45}   
Assume that $\Oomega$ is a uniquely ergodic $(r,R)$-system with invariant    
probability measure $\mu$ and   
$A\in\CalA$. Then, for any van Hove sequence $(Q_n)$ it follows that   
$$   
\lim_{n\in\NN}\frac{1}{|Q_n|}\tr (A_\omega|_{Q_n})=\tau (A)   
$$   
for every  $\omega\in\Omega$.   
\end{prop}   
Clearly,   
$A_\omega|_{Q}$ denotes the restriction of $A_\omega$ to the subspace $\ell^2(\omega\cap Q)$ of $\ell^2 (\omega)$. Note that this subspace is finite-dimensional, whenever $Q\subset\RR^d$    is bounded.   
   
We  will use here the shorthand $A_\omega(p,q)$   
for the kernel associated with $A_\omega$.    
   
\begin{proof}   
Fix a nonnegative $u\in C_c (\RR^d)$ with $\int_{\RR^d} u(x)   
dx =1$ and support contained in $B_r (0)$  and let $f(\omega,p):=u(p)$. Then   
\begin{eqnarray*}   
\tau(A)&=&\int_\Omega   
\tr(A_\omega M_u)\,d\mu(\omega)\\   
&=&\int_\Omega\left(\sum_{p\in\omega}A_\omega(p,p)u(p)\right)\,d\mu(\omega)\\   
&=&\int_\Omega F(\omega)\,d\mu(\omega) ,   
\end{eqnarray*}   
where    
$$   
F(\omega):=\sum_{p\in\omega}A_\omega(p,p)u(p)$$   
is continuous by virtue of \cite{oamp}, Proposition 2.5 (a). Therefore,   
the ergodic theorem for uniquely ergodic systems implies that for every  $\omega\in \Omega$:   
$$   
\frac{1}{| Q_n|}\int_{Q_n}F(\omega+t)dt\to\int_\Omega F(\omega)\,d\mu(\omega) .   
$$   
On the other hand,   
\begin{eqnarray*}   
\frac{1}{| Q_n|}\int_{Q_n}F(\omega+t)dt &=& \frac{1}{| Q_n|}\int_{Q_n}   
\left(\sum_{p\in\omega+t}A_{\omega+t}(p,p)u(p)\right)\, dt\\   
&=&\frac{1}{| Q_n|}\underbrace{\int_{Q_n}   
\left(\sum_{q\in\omega}A_{\omega}(q,q)u(q+t)\right)\, dt}_{I_n}   
\end{eqnarray*}   
by covariance of $A_\omega$. Since $\supp u\subset B_r(0)$ and   
the integral over $u$ equals $1$,  every $q\in\omega$ such that   
$q+B_r(0)\subset Q_n$ contributes $A_\omega(q,q)\cdot1$ in the sum under the    
integral $I_n$. For those $q\in\omega$ such that   
$q+B_r(0)\cap Q_n=\emptyset$, the corresponding summand gives $0$. Hence   
\begin{eqnarray*}   
|\frac{1}{| Q_n|}\left(\sum_{q\in\omega\cap Q_n}A_{\omega}(q,q)-I_n\right) |   
&\le &   
\frac{1}{| Q_n|}\cdot\#\{ q\in\partial_{2r}Q_n\}\cdot \| A_\omega \|\\   
&\le & C\cdot \frac{|\partial_{2r}Q_n|}{|Q_n|} \to 0   
\end{eqnarray*}   
since $(Q_n)$ is a van Hove sequence.   
\end{proof}   
   
A variant of this proposition also holds in the measurable situation. 

\begin{prop}\label{prop46}
Let $\mu$ be an ergodic measure on $\Oomega$.  Let $A\in \CalN$ and an
increasing  van Hove sequence $(Q_n)$ of compact sets  in $\RR^d$ with $\RR^d = \cup Q_n$, $0\in Q_1$ and $|Q_n - Q_n|\leq C |Q_n|$ for some $C>0$ and all $n \in \NN$ be given. Then, 
$$   
\lim_{n\in\NN}\frac{1}{|Q_n|}\tr (A_\omega|_{Q_n})=\tau (A)   
$$   
for $\mu$-almost  every  $\omega\in\Omega$. 
\end{prop}   
\begin{proof} The proof follows along similar lines as the proof of the
  preceeding proposition after replacing the ergodic theorem for
  uniquely ergodic systems by the Birkhoff ergodic theorem.  Note that for $A\in \CalN$, the function $F$
  defined there is bounded and measurable.  \end{proof}   
   
In the proof we used ideas of Hof \cite{hof2}. The following result   
finally establishes an identity that one might call an abstract   
Shubin's trace formula. It says that the abstractly defined trace $\tau$   
is determined by the integrated density of states. The latter is the   
limit of the following eigenvalue counting measures. Let, for   
selfadjoint $A\in\CalA$ and $Q\subset\RR^d$:   
$$   
\langle\rho[A_\omega,Q],\varphi\rangle := \frac{1}{| Q|}   
\tr (\varphi(A_\omega|_Q)) , \varphi\in C(\RR) .   
$$   
Its distribution function is denoted by $n[A_\omega,Q]$, i.e.
$n[A_\omega,Q](E)$ gives the number of eigenvalues below $E$ per volume   
(counting multiplicities).    
\begin{theorem}   
Let $\Oomega$ be a uniquely ergodic $(r,R)$-system and $\mu$ its ergodic   
probability measure. Then, for selfadjoint $A\in\CalA$ and any van Hove    
sequence $(Q_n)$,
$$   
\langle\rho[A_\omega,{Q_n}],\varphi\rangle\to \tau(\varphi(A))\mbox{  as  }   
n\to\infty 
 $$   
for every $\varphi\in C(\RR)$ and every  $\omega\in \Omega$. 
Consequently, the measures $\rho_\omega^{Q_n}$ converge weakly to the measure    
$\rho_A$   
defined above by $\langle\rho_A,\varphi\rangle:=\tau(\varphi(A))$, for    
every  $\omega\in  \Omega$.   
\end{theorem}   
\begin{proof}   
Let $\varphi   
\in C(\RR)$ and $(Q_n)$ be a van Hove sequence. From Proposition \ref{prop45},    
applied to $\varphi(A)=   
(\varphi(A_\omega))_{\omega\in\Omega}$, we already know that   
$$   
\lim_{n\in\NN}\frac{1}{|Q_n|}\tr (\varphi(A_\omega)|_{Q_n})=\tau (\varphi(A))   
$$   
for arbitrary  $\omega\in \Omega$. Therefore, it remains to show that   
$$   
\lim_{n\in\NN}\frac{1}{|Q_n|}\left(\tr (\varphi(A_\omega)|_{Q_n})-   
\tr (\varphi(A_\omega|_{Q_n}))\right)=0\qquad (\ast) .   
$$   
This latter property is stable under uniform limits of functions $\varphi$,   
since both $\varphi(A_\omega|_{Q_n})$ and $\varphi(A_\omega)|_{Q_n}$ are   
operators of rank dominated by $c\cdot |Q_n|$.   
   
It thus suffices to consider a polynomial $\varphi$. 

Now, for a fixed polynomial $\varphi$ with degree $N$, there exists a constant $C=
C(\varphi) $ such that
$$ \|\varphi(A) - \varphi(B)\| \leq  C \| A- B\| (\|A\| + \|B\|)^N$$
for any $A,B$ on an arbitrary Hilbert space. In particular, 
$$ \frac{1}{|Q_n|}\left|\tr (\varphi(A_\omega)|_{Q_n})-   
\tr (\varphi(B_\omega)|_{Q_n})\right| \leq C \|A_\omega -
B_\omega\|(\|A_\omega\| + \|B_\omega\|)^N$$
and
$$ \frac{1}{|Q_n|}\left|\tr (\varphi(A_\omega|_{Q_n}))-   
\tr (\varphi(B_\omega|_{Q_n}))\right| \leq C \|A_\omega -
B_\omega\|(\|A_\omega\| + \|B_\omega\|)^N$$
for all $A_\omega$ and $B_\omega$. 

Thus, it suffices to show $(\ast)$ for a polynomial $\varphi$ and $A\in
\CalAf$, as this algebra is dense in $\CalA$.  Let such $A$ and $\varphi$ be given. 

Let  $R_a$ the range of the kernel $a\in C(\xo )$ corresponding to $A$.   
   Since the kernel of $A^k$ is the $k$-fold convolution product $b:=   
a\cdots a$ one can easily verify that the range of $A^k$ is bounded   
by $N\cdot R_a$. Thus, for all $p,q\in\omega\cap Q_n$ such that the   
distance of $p,q$ to the complement of $Q_n$ is larger than $N\cdot R_a$,   
the kernels of $A^k_\omega|_{Q_n}$ and $(A|_{Q_n})^k$ agree for $k\le N$.   
We get:   
$$   
((\varphi(A_\omega)|_{Q_n})\delta_q|\delta_p)=b(p,\omega,q)   
=   
(\varphi(A_\omega|_{Q_n})\delta_q|\delta_p) .   
$$   
Since this is true outside $\{ q\in\omega\cap Q_n: \dist (q,Q_n^c)   
> N\cdot R_a\}   
\subset \partial_{N\cdot R_a}Q_n$ the matrix elements of    
$(\varphi(A_\omega)|_{Q_n})$ and $\varphi(A_\omega|_{Q_n})$ differ at   
at most $c\cdot |\partial_{N\cdot R_a}Q_n|$ sites, so that   
$$   
| \tr (\varphi(A_\omega)|_{Q_n}) -\tr (\varphi(A_\omega|_{Q_n}))|   
\le C\cdot    
|\partial_{N\cdot R_a}Q_n| .   
$$   
Since $(Q_n)$ is a van Hove sequence, this gives the desired convergence.   
   
%So far we have proven that $(\ast)$ holds for $A\in\CalAf$ and $\phi\in   
%C(\RR)$. We want to deduce the same convergence for arbitrary $A\in\CalA$.   
 %Fix $(Q_n)$ and $\varphi$ and choose a sequence $A_n$ in   
%$\CalAf$ that converges to $A$. Then we get a set of measure zero   
%outside which we have $(\ast)$ for every $A_n$. It remains to   
%show that $(\ast)$ extends to the limit $A$. Clearly, the rhs converges since   
%finite traces are continuous with respect to norm convergence.   
%To see that also the lhs converges, note that   
%$$   
%|\tr (\varphi(A_{m,\omega}|_{Q_n})) - \tr (\varphi(A_{\omega}|_{Q_n}))|   
%\le C\cdot |Q_n|\cdot \|(\varphi(A_{m,\omega}|_{Q_n}))-   
%(\varphi(A_{\omega}|_{Q_n}))\| ,   
%$$   
   
%where we estimated the lhs by the dimension times the maximal size of matrix    
%elements. This yields the required uniformity of the convergence as    
%$n\to\infty$.   
 %By choosing a countable dense subset $D$ of $C_0(\RR)$ we find a set $N$ of    
%measure $0$ such that $(\ast)$ holds   for all $\omega\not\in N$ and   
%$\varphi\in D$.  Note that $(\ast)$  is not changed if $\varphi$ is   
%changed outside $[-\| A\| ,\| A\| ]$. Therefore, for any   
%$\varphi\in C(\RR)$ we find an approximating sequence in $D$. By the   
%arguments above this yields validity of $(\ast)$ for the limit $\varphi$   
%and hence the asserted weak convergence.    
\end{proof}

The above statement has many precursors: \cite{as,be1,be2,pf,shu} in the   
context of almost periodic, random or almost random operators on   
$\ell^2(\ZZ^d)$ or $L^2(\RR^d)$. It generalizes results by Kellendonk    
\cite{ke} on tilings associated with primitive substitutions. Its proof   
relies on ideas from \cite{as,be1,be2,ke} and \cite{hof2}. Nevertheless,   
it is new in the present context.

For completeness reasons, we also state the following result.

\begin{theorem}   
Let $\Oomega$ be  an  $(r,R)$-system with an ergodic probabiltiy
measure $\mu$ . Let  $A\in\CalA$ be selfadjoint  $(Q_n)$ be an increasing  van Hove    
sequence $(Q_n)$ of compact sets in $\RR^d$ with $\cup Q_n = \RR^d$, $0\in Q_1$and $|Q_n - Q_n|\leq C | Q_n|$ for some $C>0$ and all $n\in \NN$. Then,  
$$  
\langle\rho[A_\omega,{Q_n}],\varphi\rangle\to \tau(\varphi(A))\mbox{  as  }   
n\to\infty   
$$   
for $\mu$-almost every $\omega\in \Omega$. 
Consequently, the measures $\rho_\omega^{Q_n}$ converge weakly to the measure    
$\rho_A$   
defined above by $\langle\rho_A,\varphi\rangle:=\tau(\varphi(A))$, for
$\mu$-almost     every  $\omega\in  \Omega$.   
\end{theorem}   

The {\it Proof} follows along similar lines as the proof of the previous
theorem with two modifications:  Instead of Proposition \ref{prop45}, we
use Proposition \ref{prop46};  and instead of dealing with arbitrary
polynomials we choose a countable set of polynomials which is dense in
$C_c ([-\|A\|-2, \|A\| +2])$. 

\medskip

The primary object from the    
physicists point of view is the finite volume limit:   
$$N[A](E):= \lim_{n\to\infty}n[A_\omega,Q_n](E)$$   
known as the integrated density of states. It has a striking relevance   
as the number of energy levels below $E$ per unit volume, once its   
existence and independence of $\omega$ are settled.   
    
The last two theorems  provide the mathematically rigorous    
version. Namely, the distribution    
function $N_A(E):=\rho_A(-\infty,E]$ of $\rho_A$ is the right   
choice. It gives a limit of finite volume counting measures since   
$$   
\rho[A_\omega,Q_n]\to \rho_A\mbox{  weakly as  }n\to\infty .   
$$   
Therefore, the desired independence of $\omega$ is also clear.   
Moreover,    
by standard arguments we get that the distribution functions of   
the finite volume counting functions converge to $N_A$ at points   
of continuity of the latter.    
   
In \cite{ls2} we present a much   
stronger result for uniquely ergodic minimal DDS that extends   
results for onedimensional models by the first named author,   
\cite{le2}. Namely we prove   
that the distribution functions converge uniformly, uniform in    
$\omega$. The above result can then be used to identify the   
limit as given by the tace $\tau$. Let us stress the fact that   
unlike in usual random models, the function $N_A$ does exhibit   
discontinuities in general, as explained in \cite{kls}.   
   
Let us end by emphasizing that the assumptions we posed   
are met by all the models that are usually considered in connection   
with quasicrystals. In particular, included are those Delone sets         
that are constructed by the cut-and-project method as well as models   
that come from primitive substitution tilings.

\end{document}